\documentclass[12pt,preprint]{aastex}

\shorttitle{Physics of Interpulses}

\shortauthors{S. A. Petrova}

\begin{document}
\title{Physics of Interpulse Emission in Radio Pulsars}
\author{S. A. Petrova}
\affil{Institute of Radio Astronomy, 4, Chervonopraporna Str.,
61002 Kharkov, Ukraine}

\begin{abstract}
The magnetized induced Compton scattering off the particles of the
ultrarelativistic electron-positron plasma of pulsar is
considered. The main attention is paid to the transverse regime of
the scattering, which holds in a moderately strong magnetic field.
We specifically examine the problem on induced transverse
scattering of the radio beam into the background, which takes
place in the open field line tube of a pulsar. In this case, the
radiation is predominantly scattered backwards and the scattered
component may grow considerably. Based on this effect, we for the
first time suggest a physical explanation of the interpulse
emission observed in the profiles of some pulsars. Our model can
naturally account for the peculiar spectral and polarization
properties of the interpulses. Furthermore, it implies a specific
connection of the interpulse to the main pulse, which may reveal
itself in the consistent intensity fluctuations of the components
at different timescales. Diverse observational manifestations of
this connection, including the moding behavior of PSR B1822-09,
the peculiar temporal and frequency structure of the giant
interpulses in the Crab pulsar, and the intrinsic phase
correspondence of the subpulse patterns in the main pulse and the
interpulse of PSR B1702-19, are discussed in detail. It is also
argued that the pulse-to-pulse fluctuations of the scattering
efficiency may lead to strong variability of the interpulse, which
is yet to be studied observationally. In particular, some pulsars
may exhibit transient interpulses, i.e. the scattered component
may be detectable only occasionally.
\end{abstract}

\notetoeditor{Please print this paper, or at least the Appendix,
in single column}

\section{INTRODUCTION}
The profiles of some pulsars include the component usually called
interpulse (IP), which is separated from the main pulse (MP) by
approximately a half of the pulsar period. The IPs may be as weak
as a few per cent of the MP and typically have much steeper
spectra, being most pronounced at low frequencies \citep{bu77}. In
several pulsars, the IP is connected to the MP with a barely
distinguishable emission bridge, and the low-level emission may
extend over practically the whole pulse period
\citep[e.g.][]{hc81}. In general, the IP is located not exactly
midway between the successive MPs, but may be shifted by up to a
few tens degrees toward earlier or later pulse phases. In case of
still larger shifts the IP is rather referred to as the postcursor
or the precursor to the MP \citep[e.g.][]{wei81,rr97,wiel99}. Note
that the pulse profile may contain a few such components at once.
For example, the Crab pulsar has a total of six components outside
of the MP, which exhibit pronounced frequency evolution
\citep{mh96,c04}. The emission components outside of the MP are
characteristic of the short-period pulsars, $P\la 0.6$ s
\citep{ml77}, and especially of the millisecond pulsars
\citep{wiel99}. The IPs are met in $\sim 2\%$ of the normal
pulsars and $\sim 40\%$ of the millisecond ones.

As a rule, the IP separation from the MP does not change with
frequency \citep{hf86}. This contrasts with the spectral behaviour
of the component separations inside the MP, which are known to
increase with wavelength, signifying the overall broadening of the
emission cone. Besides that, the IPs usually have distinctive
polarization characteristics, showing higher precentage of linear
polarizaton and a shallow position angle swing, the position angle
itself being not very much different from that of the MP. Although
the classical model of rotating vector explains successfully the
position angle swing across the MP in a number of pulsars, its
application to the profiles with IPs often faces difficulties
\citep[e.g.][]{rr97}.

The single-pulse observations further reveal the peculiarities of
the IP emission. The IP intensity appears modulated at various
timescales. Similarly to the MPs, the IPs may exhibit a number of
fluctuation phenomena, such as microstructure, subpulse drift,
pulse-to-pulse intensity modulation, mode changing, and giant
pulses. In PSR B0950+08, the IP shows microstructure comparable
with that in the MP \citep{hb81}. Recently it has been found that
the subpulse patten in the IP of PSR B1702-19 has the same
periodicities as that in the MP, and moreover, the two patterns
are intrinsically in phase \citep{welt07}. This striking result
directly testifies to the physical relation between the MP and IP
emissions. Some indirect manifestations of such a relation have
presumably been known previously. In particular, in PSR B0950+08,
a strong MP is followed by a strong IP \citep{hc81}, whereas in
PSR B1055-52 a strong IP is followed by a strong MP \citep{b90}.
In both cases the strong components appear separated by more than
a half of the pulse period. It is still obscure whether this
peculiar intensity correlation is an artifact of the subpulse
drift or not.

Apart from the weak intensity fluctuations of the IP emission
similar to those in the MP, occasional strong pulses, with
intensities as large as about the MP intensity, can be met in
these pulsars at the position of the IP \citep{hc81,b90}. These
transient events are so rare that neither the average IP profile
nor the above mentioned MP-IP correlation are affected.

In PSR B1822-09, the IP shows quite different fluctuation
behaviour, participating in mode changing \citep{f81,f82,g94}. The
IP appears pronounced only in the so-called quiet mode, i.e. in
the sequence of weak enough pulses. In the bright mode, a strong
precursor arises some $20^\circ$ ahead of the MP, whereas the IP
becomes markedly weaker. Thus, the IP intensity is anticorrelated
with both the MP and precursor intensities.

Provided that the giant pulse activity is characteristic of the MP
(e.g. in the Crab pulsar and PSR B1937+21), giant pulses can also
be met in the IP, though they are not necessarily simultaneous in
the two components at a given frequency. In the Crab pulsar, giant
MPs and IPs are found to exhibit quite distinct temporal and
frequency structures \citep{eh07}. Giant MPs consist of one or
several broadband microbursts made up of shorter and narrowband
nanoshots ($\delta t\sim 10/\nu $, $\delta\nu /\nu\sim 0.1$).
Giant IPs contain microsecond-long trains of proportionally spaced
emission bands ($\delta\nu /\nu\sim 0.06$), which are grouped into
regular band sets.

In summary, the IP emission is characterized by a number of
peculiarities, and at the same time shows diversiform
manifestations of its physical connection to the MP. All this
calls for theoretical explanation. In the preceding literature,
the IPs are interpreted in terms of several geometrical models.
One of the models suggests that the IP is emitted from the
magnetic pole opposite to that responsible for the MP and the
magnetic axis is nearly orthogonal to the rotational axis, in
which case the emission from both poles is alternately seen by an
observer. This two-pole model can satisfactorily explain the main
features of the IP components, except for their physical relation
to the MP emission and the continuous emission bridge, which may
connect them to the MP. These difficulties are removed in the
single-pole models. In the first version of such a model, the MP
and IP are identified with the two components of a hollow emission
cone under the condition of unusually large angular extent of the
cone and/or approximate alignment of the magnetic and rotational
axes of a pulsar \citep{ml77}. Although this model naturally
explains the emission bridge between the MP and the IP as well as
the physical connection of the components, it is not clear why the
components so much differ in intensity and their separation does
not change with frequency like that between the conal components
of the ordinary narrow profiles. Later on \citet{g85} attempted to
improve the single-pole model in order to avoid these
difficulties. It has been assumed that the two concentric hollow
cones, the inner and outer ones, are centered at the magnetic
axis, which is almost aligned with the rotational axis. Then the
portions of the inner and outer emission cones grazed by the sight
line form the MP and the IP in the resultant pulse profile.

Recently \citet{d05} have suggested a bidirectional model of
pulsar emission and applied it to the peculiar profile of PSR
B1822-09. In that model, the MP and the precursor originate
independently at different altitudes above the same magnetic pole
and the precursor emission intermittently reverses its direction
to form the IP. This so-called inward emission directed toward the
neutron star can be observable provided that the magnetic and
rotational axes are nearly orthogonal. The physics underlying the
reversal of the emission direction is not specified, but for any
conceivable switching mechanism it is difficult to explain the
connection of the emission direction to the MP intensity.

In the present paper, we suggest the first physical model of IPs.
In contrast to the above mentioned geometrical models, it is aimed
at explaining the spectral and polarization peculiarities of the
IP emission as well as the nature of the MP-IP connection.
Recently we have proposed the physical mechanism of the precursors
\citep{p07}. It is based on induced scattering of the MP emission
into the background. It has been found that in case of a
superstrong magnetic field the scattered radiation is directed
almost along the field. Then, because of rotational aberration in
the scattering region, the scattered component appears in the
pulse profile up to $30^\circ$ ahead of the MP. In the present
paper, we extend the theory of magnetized induced scattering to
the case of a moderately strong magnetic field. It will be shown
that in this approximation the MP emission is scattered in the
direction antiparallel to the ambient magnetic field and may form
the profile component roughly midway between the MPs.

The plan of the paper is as follows. In Section 2, we examine the
problem on the radio beam scattering into the background in
application to pulsars. The properties of the scattered component
are compared with the observed features of the IP emission in
Section 3. A summary of our model of the IP formation is given in
Section 4. Basic formalism of induced scattering in the
approximation of moderately strong magnetic field is given in
Appendix.

\section{TRANSVERSE SCATTERING IN PULSAR MAGNETOSPHERE}
\subsection{General Considerations}
Pulsar magnetospheres contain the ultrarelativistic
electron-positron plasma, which streams along the open magnetic
lines. The radio emission is believed to originate deep inside the
open field line tube, and therefore it should propagate through
the plasma flow. As the brightness temperatures of pulsar radio
emission are extremely high, the induced scattering off the plasma
particles may be substantial. The induced scattering of radio
emission by the non-magnetized pulsar wind has been considered by
\citet{wr78}. However, inside the magnetosphere of a pulsar, the
magnetic field may be strong enough to affect the scattering
process considerably. This happens on condition that the radio
frequency in the particle rest frame is much less than the
electron gyrofrequency, $\omega^\prime\equiv\omega\gamma
(1-\beta\cos\theta)\ll\omega_G\equiv eB/(mc)$ (here $\gamma$ is
the particle Lorentz-factor, $\beta$ the velocity in units of $c$,
$\theta$ is the angle the incident photon makes with the particle
velocity, and the quantities not denoted by primes correspond to
the laboratory frame). The magnetized scattering has been studied
in a number of papers
\citep[e.g.][]{c70,c71,hk74,bs76,bm79,ou83,c86}.

In the vicinity of the emission region of a pulsar, the regime of
magnetized scattering is certainly valid. As the magnetic field
strength decreases with distance from the neutron star, $B\propto
r^{-3}$, at high enough altitudes $\omega^\prime =\omega_G$, i.e.
the cyclotron resonance takes place. The resonance region
typically lies in the outer magnetosphere, and hence the regime of
magnetized scattering, $\omega^\prime\ll\omega_G$, holds over a
substantial part of the open field line tube well below the light
cylinder.

The pulsar radio emission is believed to be generated at the
frequencies of order of the local Lorentz-shifted proper plasma
frequency, $\omega\sim\omega_p\sqrt{\gamma}$, where
$\omega_p\equiv\sqrt{4\pi n_ee^2/m}$ and $n_e$ is the plasma
number density \citep[but see][for the criticism of this
point]{gm99}. Hence, in the emission region and its close
neighborhood, the scattering is the collective process. The
induced scattering of different types of the plasma waves is an
important ingredient of various scenarios of the pulsar emission
mechanism \citep[e.g.][]{l79,l92,l93,l96,l98}. Besides that, the
resultant radio waves may participate in the induced three-wave
interactions \citep{gk93,l98,lm06}. As the plasma number density
decreases with the distance from the neutron star, $n_e\propto
B\propto r^{-3}$, well above the emission region
$\omega\gg\omega_p\sqrt{\gamma}$, in which case the plasma effects
are negligible and the induced scattering is a single-particle
process \citep[e.g.][]{l98}. In the present paper, we dwell on the
magnetized induced scattering in the single-particle
approximation. Then the incident radiation presents the transverse
waves linearly polarized either in the plane of the ambient
magnetic field or perpendicularly to this plane.

Actually, there are two regimes of magnetized scattering, the
longitudinal and transverse ones \citep[e.g.][]{bs76,ou83}. As the
strength of the external magnetic field tends to infinity, the
excited motion of a particle in the field of the incident wave is
confined to the magnetic field line. This is a so-called
longitudinal scattering. Because of a purely longitudinal motion
of the particle in this regime, only the photon states with the
polarization in the plane of the ambient magnetic field are
involved in the scattering. In case of somewhat weaker magnetic
fields, the excited motion of a particle presents a drift in the
crossed fields, the electric field of the incident wave and the
ambient magnetic field. In case of large enough transverse
component of the perturbed particle velocity the character of the
scattering changes substantially. In particular, this so-called
transverse scattering involves the photons of both polarizations,
with the electric vectors in the plane of the magnetic field and
perpendicular to this plane. In case of spontaneous scattering of
the radiation directed at the angle $1/\gamma\ll\theta\ll 1$ to
the magnetic field, the longitudinal regime holds if
$\theta\gamma\omega^\prime/\omega_G\ll 1$, whereas the transverse
one on condition that
$(\theta\gamma)^{-1}\ll\omega^\prime/\omega_G\ll 1$
\citep[e.g.][]{ou83}. For the induced scattering of radio beam
into background, which will be studied in the present paper, the
condition of switching between the regimes is expected to be
somewhat modified, and this question will be examined in detail in
Section 2.5.

At the conditions relevant to pulsar magnetosphere, both regimes
are believed to be appropriate. Previous studies of magnetized
induced scattering in application to pulsars have concentrated on
the longitudinal regime \citep{bs76,lp96,p04a,p04b,p07}. This
process appears efficient, and it may cause a number of
observational consequences. In particular, induced longitudinal
scattering of the radio beam into the background may account for
the low-frequency turnovers in pulsar spectra \citep{lp96},
whereas the scattered component may be identified with the
precursor to the MP \citep{p07}; the scattering inside the beam
results in the photon focusing, which may underlie the formation
of microstructure in pulsar profiles \citep{p04a}; induced
scattering between the two beams of substantially different
frequencies and orientations leads to significant intensity
redistribution in frequency and can explain giant pulses along
with their nanostructure \citep{p04b}.

The induced scattering in the transverse regime has been briefly
discussed in \citet{bs76}. General formalism of this process is
developed in the Appendix of the present paper. The kinetic
equations obtained are used to solve the problem of the radio beam
scattering into the background. The main motivation of our study
is that in contrast to the longitudinal scattering regime, the
radiation is believed to be scattered backwards, in the direction
antiparallel to the particle velocity. Therefore we suggest the
induced transverse scattering as a mechanism underlying the
formation of interpulses.

\subsection{Statement of the Problem}
Let the radio beam pass through  the open field line tube of a
pulsar and be scattered off the particles of the plasma flow. The
radiation of pulsars is known to be highly directional: At any
point of the pulsar emission cone it is concentrated into a narrow
beam with the opening angle $\la 1/\gamma$, whereas the width of
the emission cone itself (which determines the observed pulse
width) is typically much larger. Far from the emission region, the
radiation propagates quasi-transversely with respect to the
ambient magnetic field, $1/\gamma\ll\theta <1$. Therefore one can
neglect the beam width and represent the incident radiation by a
single wavevector ${\bf k}$ at any point of the scattering region.
(Note that at different locations within the pulsar emission cone
the wavevector orientations somewhat differ, so that the observed
pulse has a non-zero width.)

The rate of induced scattering is known to depend on the particle
recoil, i.e. on the difference of the initial and final directions
of the photons, and therefore the scattering within the beam is of
no interest here. On the other hand, by definition the induced
scattering cannot transfer the beam photons into the states where
the photon occupation numbers are initially zero. However, the
photons can still be subject to induced scattering out of the
beam, since some background photons are expected to be always
present in space. In particular, they may result from the
spontaneous scattering of the beam, which, in contrast to the
induced one, provides the photons of any orientation. Although in
pulsar case the spontaneous scattering is very inefficient and the
scattered photons are too scanty to be detectable (see, e.g.,
eq.[13] below), they can still stimulate a substantial induced
scattering from the radio beam. The beam photons should
predominantly undergo induced scattering into the background state
${\bf k_1}$, which corresponds to the maximum scattering
probability. If the induced scattering is efficient enough, the
background radiation in this state may grow significantly and
become almost as strong as the initial radio beam.

It should be noted that in the particle rest frame the photon
frequency is approximately unchanged in the scattering act.
Therefore in the laboratory frame $\omega\eta=\omega_1\eta_1$.
Thus, we examine the problem on induced scattering between the two
photon states, ${\bf k}=(2\pi\nu /c,\theta,\phi)$ and ${\bf
k_1}=(2\pi\nu_1/c,\theta_1,\phi_1)$, one of which corresponds to
the beam and another one to the background state characterized by
the maximum scattering probability; the frequencies are related as
$\nu\eta=\nu_1\eta_1$, and the occupation number of the background
photons is initially much less than that of the beam photons,
$n_1\lll n$.

\subsection{Analytical Treatment}
The kinetic equations describing the rate of the magnetized
induced scattering are derived in Appendix (see eqs.[A7]). To
proceed further we introduce several simplifications. First of
all, as we concentrate on the transverse scattering, only the last
terms in the kinetic equations (A7) should be retained. Then these
equations can be presented as
\begin{equation}
\frac{{\rm d}n}{{\rm d}r}(i\to j)=\int{\rm d\gamma F(\gamma)}\int
\widetilde{a}g^{ij}{\rm d}\Omega_1 ,
\end{equation}
where
\begin{equation}
\widetilde{a}=\frac{2\hbar k}{mc}r_e^2n(k)n_1(k\eta
/\eta_1)\frac{\omega^4\eta^4\gamma^4}{\omega_G^4}
\frac{\eta-\eta_1}{\eta_1^3}\frac{\eta\gamma^2-1}{\beta^2\gamma^5},
\end{equation}
$i,j$ stand for the polarization states of the incident and
scattered photons, and
\begin{mathletters}
\begin{eqnarray}
g^{BB}=1+\sin^2\Delta\phi ,\\
g^{BA}=(1+\cos^2\Delta\phi)\frac{(1-\eta_1\gamma^2)^2}{\beta^2\gamma^4\eta_1^2},\\
g^{AB}=(1+\cos^2\Delta\phi)\frac{(1-\eta\gamma^2)^2}{\beta^2\gamma^4\eta^2},\\
g^{AA}=(1+\sin^2\Delta\phi)\frac{(1-\eta\gamma^2)^2}{\beta^2\gamma^4\eta^2}
\frac{(1-\eta_1\gamma^2)^2}{\beta^2\gamma^4\eta_1^2}-\frac{\sin
2\theta\sin 2\theta_1\cos\Delta\phi }{2}.
\end{eqnarray}
\end{mathletters}
One can see that the kinetic equations differ from each other only
by the factor $g^{ij}$, which is generally of order unity. (Note
the symmetry of $g^{ij}$ with respect to the initial and final
photon states). Further, a detailed form of the particle
distribution function does not play a crucial role. Therefore we
consider a monoenergetic distribution with some characteristic
Lorentz-factor of the particles.

It is convenient to replace the photon occupation numbers with the
intensities, $i_\nu=2h\nu^3n({\bf k})/c^2$ and
$i_{\nu_1}=2h\nu_1^3n_1({\bf k_1})/c^2$, and making use of their
delta-functional angular distributions to integrate the kinetic
equation over the solid angle. Then we come to the following
system of equations for the spectral intensities of the beams
$I_{\nu,\nu_1}\equiv\int i_{\nu,\nu_1}{\rm d}\Omega_1$:
\begin{eqnarray}
\frac{{\rm d}I_\nu}{{\rm d}r}=-ag^{ij}I_\nu I_{\nu_1},\nonumber \\
\frac{{\rm d}I_{\nu_1}}{{\rm d}r}=ag^{ij}I_\nu I_{\nu_1},
\end{eqnarray}
where
\begin{equation}
a=\frac{4n_er_e^2}{m\gamma^3\nu^2\theta^4}
\frac{\nu^{\prime^4}}{\nu_G^4}(\eta_1-\eta),
\end{equation}
and $n_e\equiv\int F(\gamma){\rm d}\gamma$ is the particle number
density. Obviously, the radio beam intensity $I_\nu$ decreases
because of the photon scattering to the state with $\theta_1
>\theta$ and the maximum scattering probability
corresponds to $\theta_{1_{\rm max}}=\pi$ and
$\nu_1=\nu\theta^2/4\ll\nu$. Note that in this situation the
azimuthal angle $\phi_1$ is of no interest.

The system (4) has the following solution \citep[see, e.g.,
][]{p04b}:
\begin{eqnarray}
I_\nu=\frac{I(I_\nu^{(0)}/I_{\nu_1}^{(0)})\exp
(-Iag^{ij}r)}{1+(I_\nu^{(0)}/I_{\nu_1}^{(0)})\exp
(-Iag^{ij}r)},\nonumber \\
I_{\nu_1}=\frac{I}{1+(I_\nu^{(0)}/I_{\nu_1}^{(0)})\exp
(-Iag^{ij}r)},
\end{eqnarray}
where
\begin{equation}
I\equiv I_\nu +I_{\nu_1}={\rm const}
\end{equation}
is the first integral. Thus, in our approximate consideration the
induced scattering results in the net intensity transfer of the
radio beam intensity into the background. Of course, an exact
treatment of the problem taking into account the complete rather
than approximate cross-sections would show that the total
intensity $I$ somewhat decreases, the energy being deposited to
the scattering particles, and the number of photons $n+n_1$ is
conserved instead of the intensity $n\nu+n_1\nu_1$. One can see
that as a significant fraction of photons comes to the background
state, $n_1\sim n^{(0)}\ll n$, the corresponding intensity is
$n_1\nu_1\sim n^{(0)}\nu\theta^2 <n^{(0)}\nu$. Hence, of the
original energy of the radio beam, $n^{(0)}\nu$, about
$\theta^2n^{(0)}\nu$ is transferred to the background state and
$\sim (1-\theta^2)n^{(0)}\nu$ is gained by the particles. Although
$\theta\la 1$, the background intensity grows drastically (cf.
eq.[13] below), and the intensity transfer between the two states
greatly dominates the evolution of the total intensity $I$. In our
treatment, the latter is ignored and the intensity of the
efficiently growing component is intended to be roughly comparable
with the original radio beam intensity or at least be above the
detection level.

According to equation (6), the efficiency of intensity transfer is
characterized by the quantity $\Gamma =Iag^{ij}r$. Provided that
$\Gamma\ga 1$ the background intensity grows exponentially,
$I_{\nu_1}\approx I_{\nu_1}^{(0)}\exp (\Gamma)$, but still remains
much less than the original radio beam intensity, and,
correspondingly, $I_\nu\approx I_\nu^{(0)}$. A significant part of
the beam intensity is transferred to the background on a more
stringent condition:
\begin{equation}
\xi\equiv \frac{I_{\nu_1}^{(0)}}{I_\nu^{(0)}}\exp (\Gamma)\ga 1.
\end{equation}
At still larger $\Gamma$, the background intensity increases very
weakly, slowly approaching the initial radio beam intensity,
$I_{\nu_1}\approx I_\nu^{(0)}(1-1/\xi)$, whereas the beam
intensity decreases considerably, $I_\nu\approx I_\nu^{(0)}/\xi$.
It is expected that in pulsar case $\xi\ga 1$, i.e. the scattered
component is strong enough to be observable, while the beam is not
suppressed drastically.

\subsection{Numerical Estimates}
To have a notion about the efficiency of transverse scattering in
pulsar magnetosphere let us first estimate the quantity
$\Gamma_{\rm t}=Iar$, where $a$ is given by equation (5) at
$\theta_1=\pi$ and $I\approx I_\nu^{(0)}$. The original intensity
is related to the radio luminosity of a pulsar, $L$, as
\begin{equation}
I_\nu^{(0)}=\frac{L}{\nu_0\pi r^2w^2/4}\left (
\frac{\nu}{\nu_0}\right )^{-\alpha},
\end{equation}
where $w$ is the pulse width in the angular measure, $\alpha$ the
spectral index, and it is taken that $\nu_0=4\cdot 10^8$ Hz. The
number density of the scattering particles can be expressed in
terms of the Goldreich-Julian density,
\begin{equation}
n_e=\frac{\kappa B}{Pce},
\end{equation}
where $\kappa$ is the plasma multiplicity factor, and $P$ the
pulsar period. Since $I_\nu^{(0)}\propto r^{-2}$, $n_e\propto
B\propto r^{-3}$, and $\nu_G^4\propto B^4\propto r^{-12}$, the
main contribution to the scattering depth $\Gamma_{\rm t}$ comes
from the region near the cyclotron resonance, where
$\nu\gamma\theta^2/2=\nu_G$. The radius of cyclotron resonance can
be estimated as
\begin{equation}
\frac{r_c}{r_L}=0.48\left [\frac{B_\star}{10^{12}\,{\rm G}}\left (
\frac{1\, {\rm s} }{P}\right )^3\frac{4\cdot 10^8 \,{\rm
Hz}}{\nu}\frac{10}{\gamma}\left (\frac{0.3}{\theta}\right
)^2\right ]^{1/3},
\end{equation}
where $B_\star$ is the magnetic field strength at the surface of
the neutron star and all the quantities are normalized to their
typical values. Making use of equations (9)-(11), we obtain the
scattering efficiency:
\begin{equation}
\Gamma_{\rm t}=14\frac{L}{10^{28}\,{\rm
erg\,s^{-1}}}\left(\frac{\nu}{4\cdot 10^8\, {\rm Hz}}\right
)^{-\alpha-2/3}\frac{0.1\,{\rm s}}{P}\left
(\frac{B_\star}{10^{12}\,{\rm G}}\right
)^{-1/3}\frac{\kappa}{10^3}\left (\frac{\gamma}{10}\right
)^{-5/3}\left (\frac{\theta}{0.3}\right )^{-4/3}.
\end{equation}
To conclude if this is sufficient to satisfy the inequality (8)
let us estimate the initial level of the background radiation
resulting from the spontaneous transverse scattering of the radio
beam: $I_{\nu_1}^{(0)}/I_\nu^{(0)}\sim n_e\eta r{\rm d}\sigma/{\rm
d }\Omega_1$. Substituting the scattering cross-section in the
form ${\rm d}\sigma/{\rm d }\Omega_1\sim r_e^2/(\gamma\eta_1)^2$,
which is roughly applicable at $\nu^\prime\sim\nu_G$, and taking
$\eta=\theta^2/2$ and $\eta_1=2$, we find that
\begin{equation}
\frac{I_{\nu_1}^{(0)}}{I_\nu^{(0)}}=6\cdot 10^{-13}\frac{0.1\,{\rm
s}}{P}\frac{\kappa}{10^3}\frac{B_\star}{10^{12}\,{\rm G}}\left
(\frac{\theta}{0.3}\frac{10}{\gamma}\right )^2\left
(\frac{r}{10^8\,{\rm cm}}\right )^{-2}.
\end{equation}
Given that $I_{\nu_1}^{(0)}/I_\nu^{(0)}=6\cdot 10^{-13}$,
$\Gamma_{\rm t}\ga 28$ is necessary for the background intensity
to increase appreciably as a result of the induced transverse
scattering.

Obviously, at the conditions relevant to pulsar magnetosphere,
$\Gamma_{\rm t}$ can indeed be as large as a few tens. It should
be noted that the original radio luminosity entering $\Gamma_{\rm
t}$ may noticeably exceed the average radio luminosity deduced
from observations because of pulse-to-pulse fluctuations of the
radio emission and also because of intensity suppression in the
course of radio beam scattering into the background. According to
equation (12), the scattering appears more efficient for larger
original luminosities, shorter periods, and lower frequencies.
This is in line with the observational statistics.

\subsection{Transverse vs. Longitudinal Scattering}
It is interesting to compare the efficiencies of induced
scattering into the background in the transverse and longitudinal
regimes. As can be seen from equation (A7a), for a fixed
$\theta_1$ the term corresponding to the transverse scattering
dominates on condition that
$\theta^2\theta^2_1\gamma^4\nu^{\prime^4}/\nu_G^4\gg 1$. But it is
necessary to take into account that the transverse scattering is
the strongest at $\theta_{1_{\rm max}}=\pi$, whereas the
longitudinal one peaks at $\theta_{1_{\rm max}}=1/\gamma$.
Comparing the quantity $a$ given by equation (5) at
$\theta_1=\theta_{1_{\rm max}}$ with that given by equation (9) in
\citet{p07}, we find that the maximum scattering efficiencies in
the two regimes are related as
\begin{equation}
\Gamma_{\rm
t}=\frac{\nu^{\prime^4}}{\nu_G^4}\frac{\gamma^2}{3}\Gamma_{\rm l}.
\end{equation}
Thus, at frequencies $\nu^\prime\sim\nu_G$ the transverse
scattering is much more efficient. Deeper in the magnetosphere,
however, the ambient magnetic field is much stronger, the incident
intensity and the particle number density entering $\Gamma_l$ are
larger, so that the longitudinal scattering may dominate. In
general, the processes of induced scattering in the two regimes
are expected to compete in efficiency and can both be significant.
However, if $\nu^{\prime^4}\gamma^2/(3\nu_G^4)>1$ even in the
emission region, the longitudinal scattering is inefficient at
all. Note that the emission altitude is not known accurately,
especially for the millisecond pulsars. Besides that, in the
vicinity of the emission region the angle $\theta$ between the
wavevector and the ambient magnetic field is also uncertain, since
it may be determined by refraction rather than by the
magnetosphere rotation. Hence, it is not clear whether the regime
of longitudinal scattering is the case in all pulsars over the
whole radio frequency range. As for the transverse scattering, it
occurs anyway. It is worthy to mention here that, according to
equation (11), for the parameters of the pulsars which exhibit
IPs, the characteristic altitude of transverse scattering is of
order of the light cylinder radius. One can speculate that this is
the necessary condition for the pulsar to show the IP.

\subsection{Geometrical Issues}
Now let us consider the location of the scattered radiation in the
pulse profile. For the sake of simplification it is assumed that
the emission altitude is negligible as compared to the altitude of
the scattering region, the ray geometry is dominated by the effect
of magnetosphere rotation at the point of scattering, and the
magnetic axis of a pulsar is perpendicular to the rotational axis.
The scattered component resulting from the transverse scattering
is antiparallel to the ambient magnetic field, and therefore is
expected to be identified with the IP component in the pulse
profile. The ray ${\bf k}$ emitted approximately along the
magnetic axis at $t=0$ at the point O (see Fig. 1) comes to the
point of scattering S at $t=r/c$, while the magnetic axis turns by
the angle $\Omega r/c\equiv r/r_L$. The polar angle of the point
of scattering with respect to the instantaneous magnetic axis is
$r/r_L$, and in the dipolar geometry the angle between the ambient
magnetic field vector ${\bf b}$ and the magnetic axis is
$3r/2r_L$. Hence, the angle between ${\bf k}$ and ${\bf b}$ is
$r/2r_L$. In the corotating frame, the wavevector of the scattered
radiation, ${\bf k}_s^{(c)}$, is nearly antiparallel to ${\bf b}$,
whereas in the laboratory frame it is shifted because of
aberration by the angle $r/r_L$ in the direction of rotation. As
can be seen from Fig. 1, the scattered ray ${\bf k}_{{\rm IP}}$
travels from the point of scattering to the point I for $\Delta
t_{\rm IP}=2(r/c)\cos(3r/2r_L)$ and later on reaches the observer.
The main pulse is characterized by the parallel ray ${\bf k}_{\rm
MP}$, which is emitted along the magnetic axis and points toward
the observer. The ray ${\bf k}_{\rm MP}$ originates later than
${\bf k}$ by $\Delta t=\Delta\varphi/\Omega$, where $\Delta\varphi
=\pi -3r/2r_L$ is the angle between the two instantaneous
positions of the magnetic axis, and comes to the point M for the
time $r/c$. Here it should be noted that the points I and M are
equidistant from the point O, but not from the observer. The point
I is located $\Delta r=r[1-\cos(3r/2r_L)]$ farther, the
distinction becoming significant at $r\sim r_L$. Taking into
account that the difference in pulse longitude is related to the
difference in arrival times as $\Delta\lambda =2\pi\Delta t/P$, we
find finally that
\begin{displaymath}
\lambda_{\rm MP}-\lambda_{\rm IP}=\pi-\frac{5}{2}\frac{r}{r_L}-
\frac{r}{r_L}\cos\left (\frac{3r}{2r_L}\right ),
\end{displaymath}
or, equivalently,
\begin{equation}
\lambda_{\rm IP}-\lambda_{\rm MP}=\pi+\frac{5}{2}\frac{r}{r_L}+
\frac{r}{r_L}\cos\left (\frac{3r}{2r_L}\right ).
\end{equation}
Thus, the IP lags the MP by more than $\pi$. Figure 2 shows the
separation of the two components as a function of the height of
the scattering region. For a number of pulsars, the IP positions
in the pulse profile appear compatible with our model. At the same
time, it is difficult to explain the cases when the IP lags the MP
by less than $\pi$. Note that our geometrical examination is based
solely on the effect of rotational aberration in the scattering
region. A more realistic consideration including the magnetic
field distortions close to the light cylinder and the propagation
effects on the scattered radiation would possibly release the
requirement of a nearly orthogonal rotator and allow the scattered
component to lag the MP by less than $180^\circ$ to account for
the postcursors in the pulsar profiles.

\section{DISCUSSION}
The induced Compton scattering of pulsar radio emission off the
secondary plasma particles inside the open field line tube may
play a significant role. The magnetic field of a pulsar affects
the scattering process substantially. At distances of order of the
emission altitude, the magnetic field is typically strong enough
for the scattering to occur in the longitudinal regime,
$\gamma^2\nu^{\prime^4}/\nu_G^4\ll 1$. Then the photons of the
pulsar radio beam are predominantly scattered into the state with
$\theta_{1_{\rm max}}\sim 1/\gamma$ and
$\nu_1\sim\nu\theta^2\gamma^2\gg \nu$, i.e. the scattered
radiation is almost aligned with the ambient magnetic field. As
the magnetic field strength decreases with distance from the
neutron star, at higher altitudes the longitudinal scattering
regime changes for the transverse one, which holds on condition
that $1/\gamma^2\ll\nu^{\prime^4}/\nu_G^4\ll 1$. In this regime,
the orientations of the scattered photons are mostly antiparallel
to the magnetic field, $\theta_{1_{\rm max}}=\pi$, and
$\nu_1=\nu\theta^2/4\ll\nu$. The radiation scattered in the two
regimes may from separate components in the pulse profile, which
can be identified with the precursor and the IP. Apart from
explaining the geometrical location of these components in the
pulse profile, our model suggests the MP-IP and MP-precursor
connections as well as can account for the peculiar properties of
the emission components outside of the MP.

\subsection{Comparison with Other Geometrical Models}
In the geometrical aspect, our model is akin to the bidirectional
model of \citet{d05}. In that model, the precursor and IP
originate at a certain location in the outer magnetosphere and the
pulsar is approximately an orthogonal rotator. Because of the
ultrarelativistic outstream of the plasma particles along the
field lines, any conceivable emission mechanism would generate the
outward radiation directed along the magnetic field in the
corotating frame. The inward radiation is assumed to be
antiparallel to the outward one. Thus, the orientations of the
emission components and their locations in the pulse profile are
roughly the same as in our model. Note, however, that in our case
the precursor and the IP result from the scattering in the two
distinct regimes and therefore originate at somewhat different
altitudes. Besides that, the scattering sites are restricted to
the region of the radio beam passage in the rotating
magnetosphere. Correspondingly, the positions of the precursor and
the IP in the pulse profile are tightly connected to that of the
MP.

\subsection{General Features of IPs}
\subsubsection{Spectral Behavior}
As the scattered component arises in the outer magnetosphere, its
position in the pulse profile is determined by the rotational
aberration. The scattering region lies at distances of order of
the cyclotron resonance radius, $r_c$, where
$\nu\gamma\theta^2/2\equiv\nu_G$. Taking into account that
$\omega_G\propto B\propto r^{-3}$ and $\theta\propto r$ one can
see that $r_c$ is a weak function of radio frequency,
$r_c\propto\nu^{-1/5}$, and hence the IP separation from the MP is
almost the same over a wide frequency range. This is compatible
with observations: as a rule, the IP position in the profile
appears independent of frequency \citep[e.g.][]{hf86}. Note that
in the simplified case of orthogonal rotator and purely dipolar
magnetic field considered in the present paper the rotational
aberraion causes the IP to lag the MP by more than a half of the
pulsar period. A more general treatment of the pulsar geometry
including the peculiarities close to the light cylinder, is
expected to account for smaller IP separations as well.

In the approximation considered, the induced scattering results in
a net intensity transfer from the radio beam to the background,
the total intensity of the two beams being constant. Hence, the
maximum intensity of the scattered component,
$I_{\nu_1}^{\max}(\nu_1)$, is restricted to the original intensity
of the radio beam, $I_\nu^{(0)}(\nu)$. Since the intensity is
transferred to the lower frequency, $\nu_1=\nu\theta^2/4\ll \nu$,
and the pulsar radiation has a decreasing spectrum,
$I_{\nu_1}^{\rm max}(\nu_1)$ may appear considerably less than the
MP intensity at the same frequency. In a number of cases, the IPs
are indeed much weaker than the MP. It should be noted, however,
that at a fixed frequency the MP and IP may compete in intensity
if the MP is noticeably suppressed by the scattering into still
lower frequencies.

We have examined the scattering of radiation of a fixed frequency,
but, generally speaking, it is a broadband process. Since the
scattering efficiency $\Gamma_{\rm t }$ strongly depends on
frequency, the scattering should affect the spectra of the MP and
IP emissions. According to equation (12), at lower frequencies the
scattering is more efficient and, although the intensity transfer
approaches the stage of saturation, $\xi\ga 1$, the growth of the
IP component is believed to be noticeably stronger. This trend
agrees with the observations, which testify that the IP phenomenon
is most pronounced at the decameter wavelengths, at the edge of
the observed radio frequency range \citep{bu77,bu79,b87}.
Furthermore, the scattering can markedly suppress the MP emission
at low frequencies, so that the MP spectrum may somewhat flatten.
The IPs really have steeper spectra than the MPs, with the
intensity ratio of the IP and MP markedly decreasing with
frequency \citep[e.g.][]{hf86}.

\subsubsection{Statistics of Pulsars with IPs}
The estimate of the scattering efficiency (12) implies that the
intensity transfer is more significant at larger radio
luminosities, shorter periods, and weaker magnetic field
strengths. The IPs are indeed met in the pulsars with periods
$P\la 0.6$ s \citep{ml77}. Moreover, the population of normal
short-period pulsars, $P\sim 0.1$ s, is generally characterized by
larger radio luminosities than that of the long-period ones. As
for the millisecond pulsars, their luminosities are somewhat less
\citep{wiel99}, but very short periods, $P\sim 1-10$ ms, and weak
magnetic fields, $B_\star\sim 10^8-10^9$ G, favor even larger
scattering efficiencies. Note that the pulsars with IPs are indeed
more abundant in the population of the millisecond pulsars.

\subsubsection{Polarization Properties}
Our model of IP formation as a result of induced scattering of the
MP implies peculiar polarization properties of the scattered
component. In contrast to the longitudinal scattering, when the
intensity may be efficiently transferred only between the photon
states with the ordinary polarization and the scattered component
is characterized by the complete linear polarization, the
transverse scattering involves both orthogonal polarizations and
the situation is more complicated. For different channels of the
scattering, the efficiency of intensity transfer differs by the
factor $g^{ij}$, which is generally of order unity. In case of
intense scattering (see eq.[8]), the difference in $g^{ij}$ and
$I_\nu^{(0)}$ for various channels may play a significant role,
and the intensity transfer in one of the channels may
substantially dominate that in the others, so that the scattered
component may be strongly polarized. Note that the observed IPs
are typically characterized by higher percentage of linear
polarization than the MPs \citep[e.g.][]{rr97,welt07}, with the
giant IPs showing almost complete linear polarization
\citep{eh07}.

Our model also suggests a specific behavior of the position angle
of linear polarization in the IP emission. The position angle of
the scattered radiation is determined by the orientation of the
${\bf k_1}\times {\bf b}$-plane in the scattering region and
therefore should somewhat differ from that of the MP. Besides
that, the MP and IP may be dominated by different polarization
modes, in which case the position angle of the IP is additionally
shifted by $90^\circ$. As the scattering region lies in the outer
magnetosphere, in the area covered by the radio beam the magnetic
field is almost uniform and hence the position angle should remain
practically unchanged across the IP. All this is in line with the
observational data \citep[e.g.][]{rr97,mh98,welt07}.

\subsection{Intensity Modulation at Different Timescales}
As is discussed above, the origin of IPs as a result of the MP
scattering far from the emission region can account for a number
of distinctions in the properties of the MP and IP emissions. At
the same time, our model implies a physical connection between
these emissions, which is believed to manifest itself as a
correlation of the temporal fluctuations of the MP and IP and also
as a consistency of the angular and frequency structures in the
two components. The idea of the MP-IP connection is strongly
supported by a number of observational results.

\subsubsection{Subpulse Modulation}
Recent observations of PSR B1702-19 \citep{welt07} have shown that
the subpulse pattern in the IP is characterized by exactly the
same periodicities as that in the MP. Moreover, the intensity
fluctuations appear correlated with a delay of about a half of the
pulsar period. This is just what can be expected if the MP is
partially scattered into the IP. As the subpulse pattern is
independent of frequency, the subpulse modulation in the scattered
component should repeat that in the incident radiation. The MP and
the IP seen by an observer originate at different phases of pulsar
rotation, and therefore they should arise at different phases of
subpulse drift and the intensities should be correlated with a
certain temporal delay. It is worth noting that this delay should
not exactly correspond to the longitudinal separation between the
MP and IP in the profile, since the components travel somewhat
different distances to the observer.

\subsubsection{Microstructure}
The microstructure characteristic of the MP emission is also
expected to be present in the scattered component. The
observations of PSR B0950+08 have indeed revealed the
microstructure in the IP at the timescale $\tau_{\rm
IP}=90\,\mu$s, whereas in the MP $\tau_{\rm MP}=130\,\mu$s
\citep{hc81,hb81}. In our model, the relationship between the
microstructure timescales in the two components can be estimated
as follows. The intensity is transferred between the photon states
related as $\nu (1-\beta\cos\theta)=\nu_1 (1-\beta\cos\theta_1)$.
Differentiating this at fixed frequencies yields
$\nu\theta\Delta\theta =\nu_1\Delta\theta^2$, where it is taken
that $\sin\theta_1\approx\Delta\theta_1$ for $\theta_1\approx\pi$.
As $\tau_{\rm IP}/\tau_{\rm MP}=\Delta\theta_1/\Delta\theta$ and
$\nu\theta^2/2\approx 2\nu_1$, one can find that $\tau_{\rm
IP}(\nu_1)/\tau_{\rm MP}(\nu)=2/\sqrt{\theta\Delta\theta}$. Taking
into account that the microstructure timescale evolves with
frequency, $\tau_{\rm MP}\propto\nu^{-\alpha}$, we obtain finally:
\begin{equation}
\frac{\tau_{\rm IP}(\nu_1)}{\tau_{\rm
MP}(\nu_1)}=\frac{2(\theta^2/4)^{\alpha/2}}{\sqrt{\theta\Delta\theta
(\nu_1) }}.
\end{equation}
In case of PSR B0950+08, $P=0.25$ s and $\tau_{\rm MP}=130\,\mu$s,
so that the angular scale of microstructure
$\Delta\theta=2\pi\tau/P\sim 3\cdot 10^{-3}$. Then with
$\theta=0.2$ and $\alpha =2$ we have $\tau_{\rm IP}/\tau_{\rm
MP}=0.8$, which is consistent with the observed value of about
0.7. Thus, our model can account for the difference in the
microstructure timescales of the MP and the IP.

\subsubsection{Giant IPs in the Crab Pulsar}
Recent observations of the MP and IP of the Crab pulsar
\citep{eh07} have revealed quite distinct temporal and frequency
structures of the giant pulses in these components. The giant MPs
generally present one or several broadband microbursts, which
consist of narrowband ($\delta\nu /\nu\sim 0.1$) nanoshots of a
duration $\delta t\sim 10/\nu\sim 10^{-8}-10^{-9}$ s. The giant
IPs consist of the proportionally spaced narrow emission bands of
microsecond lengths organized into several band sets, which appear
at somewhat different times and exhibit a marked drift toward
higher frequencies. Below we examine the modification of the
temporal and frequency structure of the giant MPs as a result of
induced scattering into the background and argue that the
consequent structure of the scattered component is compatible with
that really seen in the giant IPs.

Since the intensity is transferred between the frequencies
approximately related as
$\nu\theta^2/2=\widetilde{\nu}/\gamma^2=2\nu_1$, the frequency
structure of the IP at a fixed pulse longitude can be derived from
differentiating this equation:
$\Delta\nu\theta^2/2+\nu\theta\Delta\theta =2\Delta\nu_1$. As the
nanoshot width is extremely small, $\delta\theta =2\pi\delta
t/P\sim 10^{-6}-10^{-7}$ (here the period of the Crab pulsar
$P=0.03$ s), the main contribution to $\Delta\nu_1$ comes from the
first term, and we have
\begin{equation}
\frac{\Delta\nu_1}{\nu_1}\sim\frac{\Delta\nu}{\nu}\sim 0.1.
\end{equation}
Thus, the proportionally spaced emission bands  of the IP(with
$\delta\nu_1/\nu_1\approx 0.06$) can naturally be attributed to
the modification of the nanoshot bandwidths as a result of induced
scattering.

The temporal durations of the bands in the IP can be estimated by
means of equation (16). Note that because of the extremely short
durations of the original nanoshots, the scattered nanoshots are
noticeably enlarged: $\tau_{\rm IP}\propto\Delta\theta^{-1/2}$.
For $\alpha =2$, $\theta=0.5$ and $\tau_{\rm MP}(\nu_1)\sim
10^{-8}-10^{-9}$ s we have $\tau_{\rm IP}\sim 0.4-1.2\,\mu$s,
which agrees with the observational data. Keeping in mind that
$\nu\theta^2/2=2\nu_1$, one can see that at a given frequency
$\nu$ different parts of the nanoshot and the neighboring
nanoshots are scattered to somewhat different $\nu_1$, leading to
a slight drift of the scattered bands toward higher frequencies,
just as is really observed. Thus, the peculiar temporal and
frequency structure of the giant IPs can indeed be explained in
terms of induced scattering of the giant MPs.

\subsubsection{Mode Changes in PSR B1822-09}
The pulsar B1822-09 is known to exhibit a peculiar mode changing
behavior: In the bright mode, its profile contains a strong
precursor and a weak IP, whereas in the weak mode the IP is strong
and the precursor is almost absent \citep{f81,f82,g94}. Within the
framework of our model, this can be interpreted as a competition
between the processes of induced scattering of the MP in the
longitudinal and transverse regimes. The variations of the
scattering efficiencies can naturally be attributed to the
fluctuations of the number density of the scattering particles.
Furthermore, the fluctuations of the plasma number density lead to
changes in the emission altitude of the MP radiation, which may
affect the applicability of the longitudinal scattering regime.

Larger multiplicities of the plasma, $\kappa$, imply larger
emission altitudes for a given frequency, in which case the
gyrofrequency appears low enough to preclude the longitudinal
scattering even in the vicinity of the emission region. At the
same time, larger $\kappa$ favor stronger transverse scattering.
As a result, the precursor component is not formed, whereas the
intensity transfer to the IP is so efficient that the MP is
markedly suppressed. As the intensity is transferred from higher
to lower frequencies, $\nu_1=\nu\theta^2/4$, and the radio beam
originally has a decreasing spectrum, the IP remains weak as
compared to the MP at the same frequency. Thus, the weak mode is
characterized by larger plasma multiplicities and,
correspondingly, by an efficient intensity transfer from higher to
lower frequencies, in which case the total intensity of the pulse
profile markedly decreases.

In the bright mode, the plasma number density is smaller, the
emission altitude lower, and the longitudinal scattering holds in
addition to the transverse one. The longitudinal scattering gives
rise to the precursor in the pulse profile, whereas the transverse
scattering, being less efficient because of lower $\kappa$, forms
a weaker IP. As is evident from equation (14), the transverse
scattering may be much more efficient in taking the intensity from
the MP at a given frequency. At the same time, if one compare the
intensity growth of the precursor and the IP at a given frequency,
it is necessary to keep in mind that these components are fed by
the MP intensity at substantially different frequencies,
$\nu_1^{(\rm Pr)}=\nu\theta^2\gamma^2\gg\nu$ and $\nu_1^{(\rm
IP)}=\nu\theta^2/4\ll\nu$. With the decreasing spectrum of pulsar
radiation, this implies much larger $\Gamma_{\rm l}/\Gamma_{\rm
t}$ than that given by equation (14), so that both scattering
efficiencies can be concurrently substantial. Since the
longitudinal scattering transfers the intensity to the precursor
from much lower frequencies, this component may be strong enough
as compared to the MP at the same frequency. Note that the
precursor in the profile of PSR B1822-09 is really comparable in
intensity to the MP and is much stronger than the IP, providing a
strong support to our scenario of intensity transfer between
widely spaced frequencies in the course of induced scattering of
the MP into the background.

\section{CONCLUSIONS}
We have considered the induced Compton scattering of radio
emission by the particles of the hot magnetized electron-positron
plasma, which streams along the open field lines of the pulsar
magnetosphere. In the presence of a strong magnetic field,
$\nu^\prime\ll\nu_G$, two scattering regimes are possible, the
longitudinal and transverse ones, both being relevant to the
pulsar case. The earlier studies of the magnetized induced
scattering in pulsars were typically restricted to the
longitudinal regime. In the present paper, we have generalized the
kinetic equation to include the transverse scattering regime as
well and have used it to solve the problem on the transverse
induced scattering of the radio beam into the background.

Our problem can be reduced to examining the scattering between the
two photon states, one of which represents the radio beam, whereas
another one is the background state corresponding to the maximum
scattering probability for the beam photons. Since in the particle
rest frame the photon frequency is almost unchanged in the
scattering act, in the laboratory frame the photon states
interacting via induced scattering obey the condition $\nu
(1-\beta\cos\theta)=\nu_1 (1-\beta\cos\theta_1)$. In contrast to
the longitudinal regime, in which case the radio beam photons are
predominantly scattered into the direction nearly along the
ambient magnetic field ($\theta_{1_{\rm max}}\sim 1/\gamma$ and
$\nu_1\sim\nu\theta^2\gamma^2\gg\nu$), the photons scattered in
the transverse regime are mostly antiparallel to the field
($\theta_{1_{\rm max}}\approx\pi$ and
$\nu_1\sim\nu\theta_1^2\ll\nu$). In both regimes, the induced
scattering results in the intensity redistribution between the two
photon states, with the total intensity remaining approximately
unchanged. This specific character of intensity evolution is
determined by the role of the magnetic field in the scattering
process and differs substantially from the non-magnetic case, when
the scattering results in the photon drift toward lower
frequencies.

The magnetized induced scattering of the radio beam into the
background, which takes place in the open field line tube of a
pulsar, is suggested to underlie the formation of additional
components in the pulse profile outside of the MP. Given that the
pulsar is a nearly orthogonal rotator, the longitudinal scattering
gives rise to the precursor component located a few tens degrees
ahead of the MP, whereas the transverse scattering forms the IP,
which lags the MP by more than $180^\circ$. The numerical
estimates show that at the conditions relevant to pulsar
magnetospheres the intensity transfer from the radio beam to the
background can indeed be efficient. Stronger scattering is favored
by shorter pulse periods, larger radio luminosities and lower
frequencies and is expected to be especially efficient in the
millisecond pulsars. All this is in line with the trends known for
the observed IP emission. Note that the IP component is fed by the
MP radiation of much higher frequencies,
$\nu_1\approx\nu\theta^2/4$. With the decreasing spectrum of the
original MP, this implies that the IP cannot be as large as the MP
at the same frequency unless the latter is substantially
suppressed by the scattering to still lower frequencies.

Our model can account for the peculiar properties of the IP
emission. As the region of efficient transverse scattering lies
far from the emission region, at distances of order of the
cyclotron resonance radius, and the characteristic scattering
altitude is an extremely weak function of frequency, the MP-IP
separation should remain almost the same over the observed radio
frequency range. Note that the scattering at somewhat lower
altitudes may also be noticeable, giving rise to the emission
bridge between the MP and the IP.

The position angle of linear polarization of the scattered
radiation is determined by the orientation of the ${\bf
k_1}\times{\bf b}$-plane in the scattering region and generally
differs from the position angle of the incident radiation.
Furthermore, the magnetic field is believed to be almost uniform
throughout the scattering region, so that the position angle swing
in the IP should be shallow. Although both orthogonal polarization
modes can grow as a result of the transverse scattering, the
efficiencies of intensity transfer may differ markedly, making
expect high percentage of linear polarization in the resultant IP
emission.

The IP formation as a result of the MP scattering implies a
physical connection between these components, which is believed to
manifest itself in their intensity fluctuations at different
timescales. Our model is strongly supported by the recent
observations of PSR B1702-19 \citep{welt07}, which have revealed
that the subpulse patterns in the MP and the IP  are intrinsically
in phase. Furthermore, the peculiar temporal and frequency
structure of the giant pulses in the IP of the Crab pulsar
\citep{eh07} may well be interpreted as a modification of the
giant MP structure in the scattering process. The peculiar moding
behavior of PSR B1822-09 is suggested to result from the
fluctuations of the physical conditions in the scattering region,
which affect the relative efficiency of the longitudinal and
transverse scatterings and lead to the observed interplay between
the resultant precursor and IP components.

The pulse-to-pulse fluctuations of the plasma number density and
the original MP intensity, which affect the scattering efficiency,
are believed to result in the peculiar intensity statistics of the
IP. Moreover, in some pulsars this component may be seen only
occasionally. Note that the transient IPs are difficult to detect
and they are yet to be found observationally, though some evidence
for occasional strong events at the IP longitudes is already known
\citep{hc81,b90}. Despite the relative weakness of the IPs in most
of the known cases, the present progress in the observational
facilities seems to promise an opportunity of the comprehensive
study of the IP emission at a single-pulse level.

In the present paper, we for the first time suggest a physical
model of the IP formation. It is believed to stimulate further
observational investigations aimed at revealing the peculiar
properties of the IP emission and the manifestations of its
connection to the MP.

\acknowledgments

I am grateful to the anonymous referee for useful suggestions and
criticisms.

\appendix
\section{BASIC FORMALISM OF THE MAGNETIZED INDUCED SCATTERING}
Following \citet{bs76,p07}, let us consider the induced scattering
in the laboratory frame between the two photon states, $\bf k$ and
${\bf k_1}$, involving the electrons with the initial momenta $p$
and $p+\delta p$. The electrons are confined to move along the
magnetic field line, and in the scattering act the momentum
parallel to the field is conserved, so that
\begin{equation}
\delta p=\hbar k\cos\theta -\hbar k_1\cos\theta_1,
\end{equation}
where $\theta$ and $\theta_1$ are the wavevector tilts to the
magnetic field for the photons in the states $\bf k$ and $\bf
k_1$. The rate of change of the photon occupation number as a
result of induced scattering is
\begin{equation}
\frac{{\rm d}n}{{\rm d}t}\frac{{\rm d}^3\bf k}{(2\pi)^3}=\int
[f(p+\delta p)-f(p)]\frac{{\rm d}P}{{\rm d}t}n_1{\rm d}p.
\end{equation}
Here $n$ and $n_1$ are the photon occupation numbers in the states
$\bf k$ and $\bf k_1$, respectively, $f(p)$ is the electron
distribution function, ${\rm d}P/{\rm d}t$ is the probability of
generating spontaneously scattered photons per electron per unit
time,
\begin{equation}
\frac{{\rm d}P}{{\rm d}t}=n\frac{\rm d^3{\bf
k}}{(2\pi^3)}\eta\frac{{\rm d}\sigma}{{\rm
d}\Omega_1}c^4\frac{{\rm d}^3\bf k_1}{\omega_1^2}\delta \left
(\omega_1-\eta\omega/\eta_1\right ),
\end{equation}
where $\eta\equiv 1-\beta\cos\theta$, $\eta_1\equiv
1-\beta\cos\theta_1$, ${\rm d}\sigma/{\rm d }\Omega_1$ is the
scattering cross-section per unit solid angle, and the argument of
the delta-function means that in the particle rest frame the
initial and final frequencies are equal,
$\omega\gamma\eta\equiv\omega_1\gamma\eta_1$. We assume that the
photon occupation numbers change only because of the photon
propagation through the stationary flow, so that ${\rm d}n/{\rm
d}t=c{\rm d}n/{\rm d}r$. Taking into account that $f(p+\delta
p)-f(p)\approx\delta p(\partial f/\partial p)$, we integrate
equation (A2) by parts, substitute equations (A1) and (A3) and
integrate this over $k_1$ with the help of delta-function to
obtain
\begin{equation}
\frac{{\rm d}n}{{\rm d}r}=\int {\rm d}\gamma
F\frac{\beta\hbar\omega}{mc^2}\int
(\cos\theta_1-\cos\theta)\frac{\partial}{\partial\gamma}\left
[nn_1\frac{{\rm d}\sigma}{{\rm d}\Omega_1}\right ]{\rm d}\Omega_1.
\end{equation}
Here $F(\gamma)$ stands for the distribution function in
Lorentz-factor.

The above kinetic equation describes the photon transfer as a
result of induced scattering in a hot magnetized plasma. We
consider the scattering of the ordinary and extraordinary
transverse electromagnetic waves. Their polarization states, with
the electric vectors in the plane of the wavevector and the
ambient magnetic field and orthogonal to this plane, are
designated as A- and B-polarizations, respectively. The classical
scattering cross-section in an arbitrary magnetic field has been
derived by \citet{c71}. Expanding their result in a power series
of $\omega^{\prime 2}/\omega_G^2$ and retaining the first two
terms yield
\begin{mathletters}
\begin{eqnarray}
\frac{{\rm d}\sigma^{AA}}{{\rm d}\Omega_1^\prime}\approx
r_e^2\sin^2\theta^\prime\sin^2\theta_1^\prime+r_e^2\frac{\omega^{\prime
2}}{\omega_G^2}\left\{\cos^2\theta^\prime\cos^2\theta_1^\prime\left
[\sin^2\Delta\phi^\prime +\frac{\omega^{\prime
2}}{\omega_G^2}(1+\sin^2\Delta\phi^\prime)\right ]\right.\nonumber
\\ \left. -\frac{1}{2}\left
(1+\frac{\omega^{\prime 2}}{\omega_G^2}\right )\sin
2\theta^\prime\sin 2\theta_1^\prime\cos\Delta\phi^\prime\right\}.
\end{eqnarray}
\begin{equation}
\frac{{\rm d}\sigma^{BA}}{{\rm d}\Omega_1^\prime}\approx
r_e^2\frac{\omega^{\prime 2}}{\omega_G^2}\cos^2\theta_1^\prime
\left [\cos^2\Delta\phi^\prime +\frac{\omega^{\prime
2}}{\omega_G^2}(1+\cos^2\Delta\phi^\prime)\right ],
\end{equation}
\begin{equation}
\frac{{\rm d}\sigma^{AB}}{{\rm d}\Omega_1^\prime}\approx
r_e^2\frac{\omega^{\prime 2}}{\omega_G^2}\cos^2\theta^\prime \left
[\cos^2\Delta\phi^\prime +\frac{\omega^{\prime
2}}{\omega_G^2}(1+\cos^2\Delta\phi^\prime)\right ].
\end{equation}
\begin{equation}
\frac{{\rm d}\sigma^{BB}}{{\rm d}\Omega_1^\prime}\approx
r_e^2\frac{\omega^{\prime
2}}{\omega_G^2}\left\{\sin^2\Delta\phi^\prime
+\frac{\omega^{\prime
2}}{\omega_G^2}[1+\sin^2\Delta\phi^\prime]\right\}.
\end{equation}
\end{mathletters}
Here the superscripts of $\sigma$ denote the initial and final
polarization states of a photon, the primes mark the quantities in
the electron rest frame, $r_e$ is the classical electron radius,
$\Delta\phi^\prime =\phi^\prime -\phi_1^\prime$,
$(\theta^\prime,\phi^\prime)$ and
$(\theta_1^\prime,\phi_1^\prime)$ are the spherical angles of the
initial and final photon wavevectors ${\bf k^\prime}$ and $\bf
k_1^\prime$ in the coordinate system with the polar axis along the
magnetic field, and $\omega^\prime =\omega_1^\prime$.

Using relativistic transformations,
\begin{equation}
\omega^\prime =\omega\gamma\eta , \quad \Delta\phi^\prime
=\Delta\phi , \quad {\rm d}\Omega_1^\prime =\frac{{\rm
d}\Omega_1}{\gamma^2\eta_1^2},\quad
\cos\theta^\prime=\frac{\cos\theta
-\beta}{1-\beta\cos\theta},\quad
\sin\theta^\prime=\frac{\sin\theta}{\gamma\eta},
\end{equation}
one can express the cross-sections in terms of the quantities of
the laboratory frame, substitute this into equation (A4) and
perform differentiation with respect to $\gamma$. It should be
noted that the dominant term of the cross-sections,
$\propto\omega^{\prime^2}/\omega_G^2$ as well as the remaining
factor in the expression under differentiation in equation (A4)
depend on $\gamma$ only implicitly, via $\eta (\beta)$ and $\eta_1
(\beta)$, which are weak functions of $\gamma$: ${\rm d}\beta/{\rm
d}\gamma=1/\beta\gamma^3\ll 1/\gamma$. Therefore it is necessary
to retain the second-order terms,
$\propto\omega^{\prime^4}/\omega_G^4$, which introduce the
explicit dependence on $\gamma$. Although they are small, their
derivatives may contribute significantly. Keeping in mind these
considerations, one can obtain the kinetic equations in the
following form:
\begin{mathletters}
\begin{eqnarray}
\frac{{\rm d}n}{{\rm d}r}(A\to A)=\frac{\hbar nr_e^2}{mc}\int {\rm
d }\gamma
F\int\frac{\sin^2\theta\sin^2\theta_1}{\gamma^6\eta^3\eta_1^3}
\left\{\frac{(\eta-\eta_1)^2}{\beta^2\gamma^3\eta_1^2}\frac{\partial
n_1k_1^2 }{\partial k_1}\right.\nonumber \\ \left.
+\frac{6(\eta_1-\eta)n_1k\eta^2}{\beta^2\gamma\eta_1^2}\left
[1-\frac{\eta+\eta_1}{2\gamma^2\eta\eta_1}\right ] \right\}{\rm
d}\Omega_1\nonumber\\ +\frac{\hbar nr_e^2}{mc}\int {\rm d }\gamma
F\int\frac{\omega^2}{\omega_G^2}\sin^2\Delta\phi\left
\{\frac{(\eta-\eta_1)^2}{\beta^2\gamma^3\eta_1^2}\frac{1}{k}
\frac{\partial n_1k_1^3}{\partial
k_1}\frac{(1-\eta\gamma^2)^2}{\beta^2\gamma^4\eta^2}
\frac{(1-\eta_1\gamma^2)^2}{\beta^2\gamma^4\eta_1^2}\right.\nonumber
\\ \left.
+\frac{2(\eta-\eta_1)kn_1\eta^3}{\beta^6\gamma^{11}\eta_1^3}\left
[\frac{(\eta\gamma^2-1)[\eta\gamma^2(2-\eta)-1]}{\eta^3}
\frac{(1-\eta_1\gamma^2)^2}{\eta_1^2 }\right.\right.\nonumber \\
\left.\left.
+\frac{(\eta_1\gamma^2-1)[\eta_1\gamma^2(2-\eta_1)-1]}{\eta_1^3}
\frac{(1-\eta\gamma^2)^2}{\eta^2}\right ]\right\}{\rm
d}\Omega_1\nonumber \\ -\frac{\hbar nr_e^2}{mc}\int {\rm d }\gamma
F\int\frac{\omega^2}{\omega_G^2}\frac{\cos\Delta\phi\sin
2\theta\sin 2\theta_1}{\mu\mu_1\beta^2\gamma^3}\left
\{\frac{(\eta-\eta_1)^2}{2\gamma^2\eta_1^2}\frac{1}{k}
\frac{\partial n_1k_1^3}{\partial k_1}\right.\nonumber
\\ \left.
-\frac{(\eta-\eta_1)kn_1\eta^3}{\eta_1^3}\left [1-\frac{\mu +\mu_1
}{2\gamma^2\mu\mu_1}\right ]\right\}{\rm d}\Omega_1\nonumber
\\ +\frac{\hbar nr_e^2}{mc}\int{\rm d}\gamma F\int\frac{\omega^4}{\omega_G^4}
\frac{(\eta
-\eta_1)n_1k\eta^4(\eta\gamma^2-1)}{\eta_1^3\beta^2\gamma} \left
[2(1+\sin^2\Delta\phi)\frac{(1-\eta\gamma^2)^2}
{\beta^2\gamma^4\eta^2}\frac{(1-\eta_1\gamma^2)^2}{\beta^2\gamma^4\eta_1^2}\right.\nonumber
\\ \left. -\cos\Delta\phi\sin 2\theta\sin 2\theta_1 \right ]{\rm
d}\Omega_1 .
\end{eqnarray}
\begin{eqnarray}
\frac{{\rm d}n}{{\rm d}r}(B\to A)=\frac{\hbar nr_e^2}{mc}\int {\rm
d }\gamma F\int\frac{\omega^2}{\omega_G^2}\cos^2\Delta\phi \left
\{\frac{(\eta-\eta_1)^2}{\beta^2\gamma^3\eta_1^2}\frac{1}{k}
\frac{\partial n_1k_1^3}{\partial
k_1}\frac{(1-\eta_1\gamma^2)^2}{\beta^2\gamma^4\eta_1^2}
\right.\nonumber
\\\left.
+\frac{(\eta-\eta_1)kn_1\eta^3}{\eta_1^3}\frac{2(\eta_1\gamma^2-1)}
{\beta^4\gamma^7\eta_1^3}[\eta_1\gamma^2(2-\eta_1)-1]\right \}{\rm
d}\Omega_1 \nonumber \\ +\frac{2\hbar nr_e^2}{mc}\int {\rm d
}\gamma
F\int\frac{\omega^4}{\omega_G^4}(1+\cos^2\Delta\phi)\frac{(1-\eta_1\gamma^2)^2}
{\beta^2\gamma^4\eta_1^2}\frac{(\eta
-\eta_1)n_1k\eta^4(\eta\gamma^2-1)}{\eta_1^3\beta^2\gamma}{\rm
d}\Omega_1,\\ \frac{{\rm d}n}{{\rm d}r}(A\to B)=\frac{\hbar
nr_e^2}{mc}\int {\rm d }\gamma
F\int\frac{\omega^2}{\omega_G^2}\cos^2\Delta\phi \left
\{\frac{(\eta-\eta_1)^2}{\beta^2\gamma^3\eta_1^2}\frac{1}{k}
\frac{\partial n_1k_1^3}{\partial
k_1}\frac{(1-\eta\gamma^2)^2}{\beta^2\gamma^4\eta^2}
\right.\nonumber
\\\left.
+\frac{(\eta-\eta_1)kn_1\eta^3}{\eta_1^3}\frac{2(\eta\gamma^2-1)}
{\beta^4\gamma^7\eta^3}[\eta\gamma^2(2-\eta)-1]\right \}{\rm
d}\Omega_1 \nonumber
\\ +\frac{2\hbar nr_e^2}{mc}\int {\rm d }\gamma
F\int\frac{\omega^4}{\omega_G^4}(1+\cos^2\Delta\phi)\frac{(1-\eta\gamma^2)^2}
{\beta^2\gamma^4\eta^2}\frac{(\eta
-\eta_1)n_1k\eta^4(\eta\gamma^2-1)}{\eta_1^3\beta^2\gamma}{\rm
d}\Omega_1.
\end{eqnarray}
\begin{eqnarray}
\frac{{\rm d}n}{{\rm d}r}(B\to B)=\frac{\hbar nr_e^2}{mc}\int {\rm
d}\gamma F\int\frac{\omega^2}{\omega_G^2} \sin^2\Delta\phi
\frac{(\eta_1-\eta)^2}{\eta_1^2\beta^2\gamma^3}\frac{1}{k}
\frac{\partial n_1k_1^3}{\partial k_1}{\rm d}\Omega_1\nonumber \\
+\frac{2\hbar nr_e^2}{mc}\int {\rm d}\gamma
F\int\frac{\omega^4}{\omega_G^4}
(1+\sin^2\Delta\phi)\frac{\eta\gamma^2-1}{\beta^2\gamma}
\frac{\eta-\eta_1}{\eta_1^3}\eta^4n_1k{\rm d}\Omega_1.
\end{eqnarray}
\end{mathletters}
The first term of the kinetic equation (A7d) is determined by the
photon spectrum and is qualitatively similar to the right-hand
side of the kinetic equation for the non-magnetic case (cf., e.g.,
eq.[2.13] in \citet{lp96}), signifying the monotonic shift of the
photon distribution toward lower frequencies in the course of the
scattering. The second term means the redistribution of photons
between the states which satisfy the condition
$\omega\eta\equiv\omega_1\eta_1$. Because of the factor $\eta
-\eta_1$, the photon occupation numbers decrease as a result of
the photon transfer to the states with $\theta_1>\theta$ and
increase on account of the photons coming from the states with
$\theta_1 <\theta$. One can find that the ratio of the second term
in equation (A7d) to the first one is
$\sim\omega^2\eta^2\gamma^2/(\omega_G^2)\chi^2\gamma^2$, where
$\chi\equiv \min (\theta,\theta_1)$. Thus, in the case of
interest, in a moderately strong magnetic field, the second term
dominates. Although the kinetic equations (A7b) and (A7c) are
somewhat more complicated, their second terms also dominate on the
same condition.

The equation (A7a) is worthy to be analyzed in more detail. Its
first term, corresponding to the first term of the cross-section
(A5a), does not contain the gyrofrequency and describes the
longitudinal scattering, which remains efficient at $B\to\infty$.
The second item of this term differs from the first one by the
factor $(\theta_1\gamma)^2$ and hence dominates at $\theta_1\gg
1/\gamma$. In the regime of transverse scattering,
$\chi^2\gamma^2\omega^{\prime 2}/\omega_G^2\gg 1$, the last term
of equation (A7a) dominates, being at least a factor of
$\chi^4\gamma^4\omega^{\prime 4}/\omega_G^4$ larger than the first
one and $\chi^2\gamma^2\omega^{\prime 2}/\omega_G^2$ larger than
the second and the third ones. The sign  of the integrand in the
last term is determined by that of $(\eta -\eta_1)$, that is the
photons are transferred to the states with $\theta_1>\theta$. This
is similar to the corresponding terms in the kinetic equations
(A7b), (A7c), and (A7d) and contrasts with the longitudinal
regime, in which case the photons are transferred closer to the
magnetic field direction, $\theta_1<\theta$ (cf. the second item
in the braces of the first term in equation (A7a); for more
details see \citet{p07}).

\clearpage

\begin{figure}
\epsscale{.9} \plotone{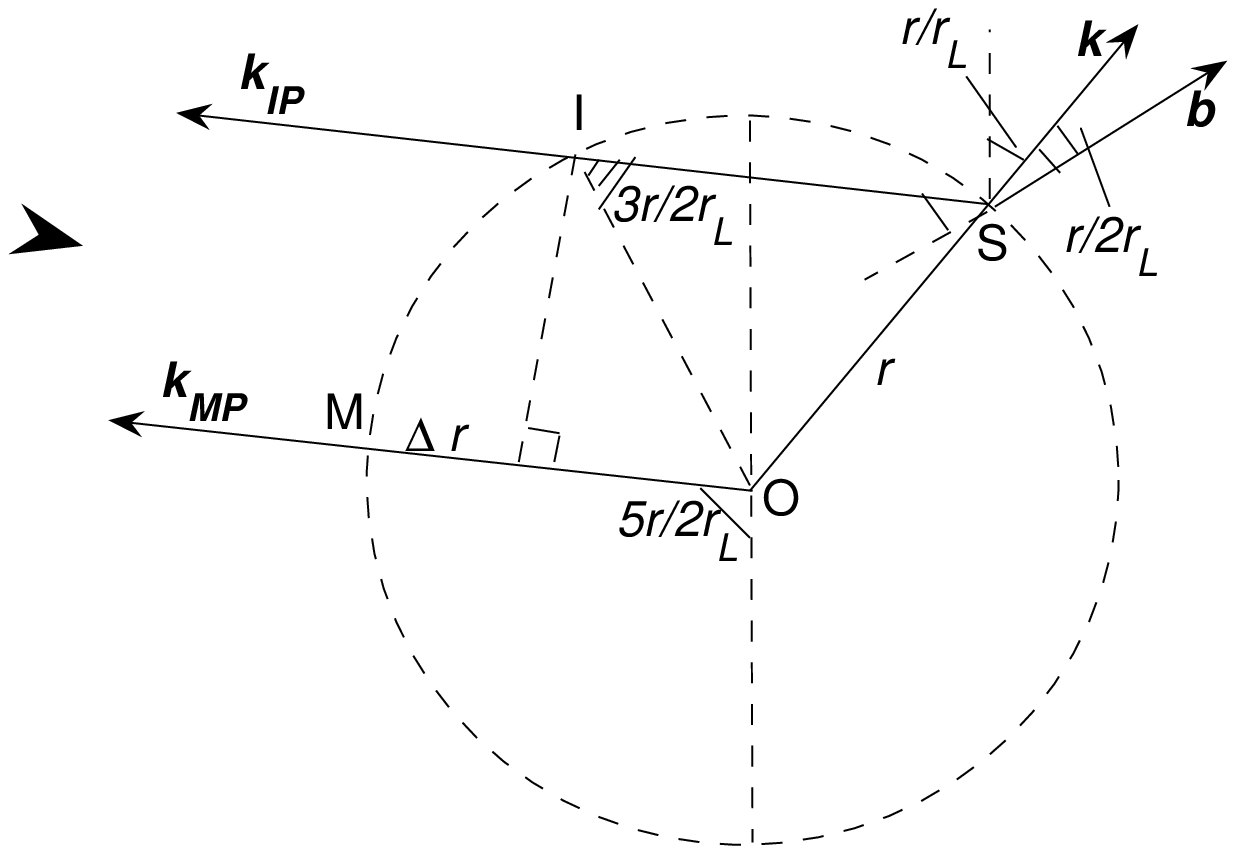} \caption{Geometrical scheme of the
scattering in the transverse regime. The pulsar is assumed to be
an orthogonal rotator with the rotational axis perpendicular to
the plane of the figure and the magnetic axis rotating
counterclockwise. For more details see the text.}
\end{figure}

\begin{figure}
\epsscale{.9} \plotone{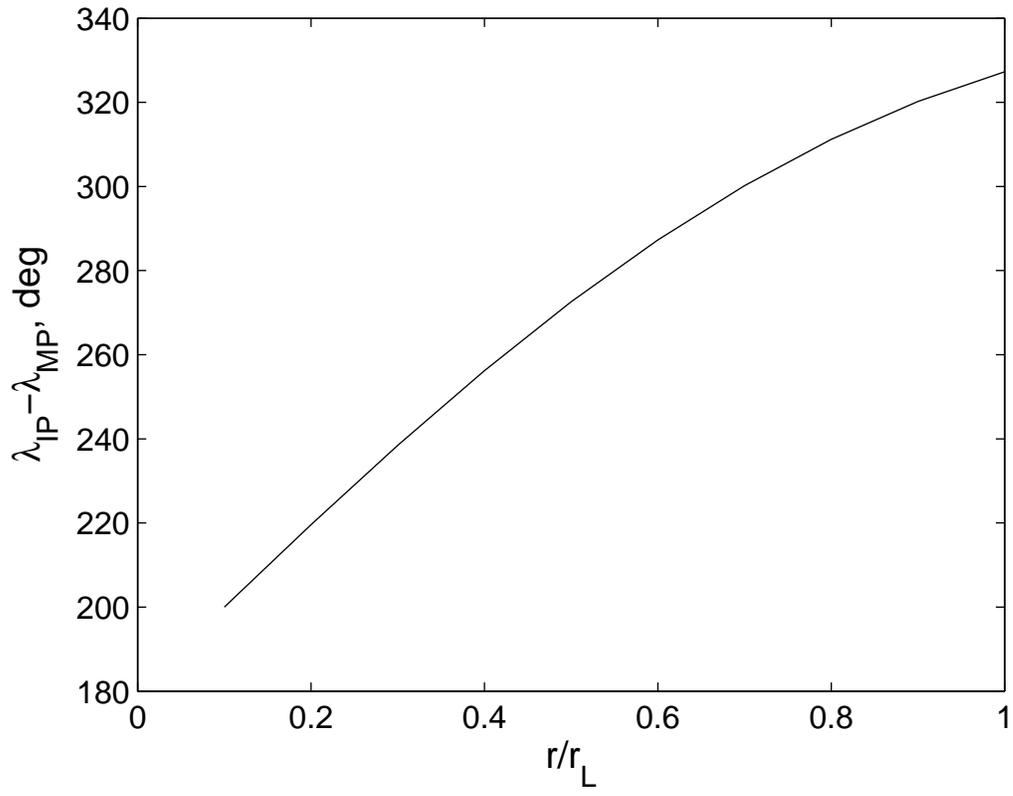} \caption{The IP location in the
pulse profile as a function of altitude of the scattering region.
}
\end{figure}

\end{document}